\documentclass[sigconf]{acmart}
\usepackage{graphicx}
\usepackage{amsmath}

\usepackage{amssymb}
\usepackage{amsthm}
\usepackage{booktabs}
\usepackage{algorithmic}
\usepackage{algorithm}
\usepackage{enumitem}

\newtheorem{remark}{Remark}[section]
\newtheorem{proposition}{Proposition}[section]

\copyrightyear{2026}
\acmYear{2026}
\setcopyright{cc}
\setcctype{by}
\acmConference[ICAIF '26]{7th ACM International Conference on AI in Finance}{November 14--17, 2026}{Milan, AA, Italy}
\acmBooktitle{7th ACM International Conference on AI in Finance (ICAIF '26), November 14--17, 2026, Milan, AA, Italy}
\acmDOI{10.1145/3858063.3858743}
\acmISBN{979-8-4007-3087-0/2026/11}

\begin{document}
\title{Diversification Across Predictors: Heterogeneous Ensemble Learning for Cross-Sectional Return Prediction}

\author{Danning Li}
\orcid{0009-0002-1626-5380}
\affiliation{%
  \institution{The Hong Kong University of Science and Technology (Guangzhou)}
  \department{Thrust of Financial Technology}
  \city{Guangzhou}
  \country{China}}
\email{dli381@connect.hkust-gz.edu.cn}

\renewcommand{\shortauthors}{Danning Li}

\begin{abstract}
Cross-sectional stock return prediction remains challenging due to extremely low signal-to-noise ratios, high-dimensional factor spaces, and non-stationary predictor effectiveness. Individual model classes exhibit complementary failure modes in this regime: neural networks overfit noise due to high variance, while linear models underfit nonlinear structure due to high bias. We propose a heterogeneous ensemble framework that extends the financial principle of diversification to the predictor level. Through a bagging procedure combining linear predictors with nonlinear predictors, each model functions as a distinct specialist whose prediction errors exhibit low cross-covariance, enabling error cancellation upon aggregation. We formalize this mechanism via bias-variance-covariance decomposition, establishing an exact mathematical correspondence between portfolio diversification and predictor aggregation. On CSI300 daily-frequency Alpha158 factor datasets, the proposed ensemble achieves a Rank IC improvement exceeding 13\% over standalone neural networks and surpasses tree-based state-of-the-art methods including XGBoost and Double Ensemble, as measured by both predictive accuracy and portfolio-based evaluation metrics.
\end{abstract}



\maketitle

\section{Introduction}

Predicting cross-sectional equity returns from quantitative factors remains one of the most challenging tasks in empirical finance.
The fundamental difficulty lies in the extremely low signal-to-noise ratio (SNR) inherent to daily-frequency return data, where the predictable component constitutes only a marginal fraction of total variation~\citep{cochrane2011presidential,welch2008comprehensive,doi:https://doi.org/10.1002/9780470061602.eqf15018}.
This regime imposes a structural dilemma on any single predictive model.
Linear models, while statistically stable, lack the representational capacity to capture nonlinear interactions among factors and thus suffer from persistent underfitting~\citep{fama1993common}.
Neural networks, conversely, possess sufficient expressiveness but are prone to fitting spurious patterns in noise, resulting in severe overfitting and poor out-of-sample generalization~\citep{10.1145/3446776}.
Gradient-boosted tree ensembles such as XGBoost~\citep{10.1145/2939672.2939785} and their variants~\citep{9338413} currently dominate empirical benchmarks, yet they remain essentially a single function class refined through sequential error correction, offering limited mechanistic diversity.
The prevailing paradigm, therefore, is to engineer an ever more powerful individual model.
We argue that this trajectory overlooks a more principled alternative: organizing a collective of complementary, moderately capable predictors.

The intellectual foundation for this alternative is already embedded in finance itself.
Markowitz's portfolio theory~\citep{e5a1bb8f-41b7-35c6-95cd-8b366d3e99bc} establishes that combining assets with imperfectly correlated returns reduces aggregate risk without proportionally sacrificing expected return.
We observe that this principle admits a direct and rigorous transposition to the space of predictive errors.
When $N$ predictors produce forecasts whose errors are weakly correlated, the variance of the averaged forecast decays at a rate of $\mathcal{O}(1/N)$, mirroring the diversification benefit of an equally weighted portfolio~\citep{breiman1996bagging, geman1992neural}.
Crucially, this variance reduction is achievable even when no individual predictor is accurate in isolation, provided that their errors are sufficiently decorrelated.
This insight resonates with the theory of collective intelligence~\citep{surowiecki2005wisdom, page2008difference}, which demonstrates that a group of diverse and independent agents, each possessing only marginal competence, can collectively outperform the best individual member.
The conditions under which such wisdom emerges, namely diversity, independence, and appropriate aggregation, are naturally satisfiable in financial prediction through architectural heterogeneity and stochastic training protocols.

\begin{figure*}[!t]
\centering
\includegraphics[width=0.8\textwidth]{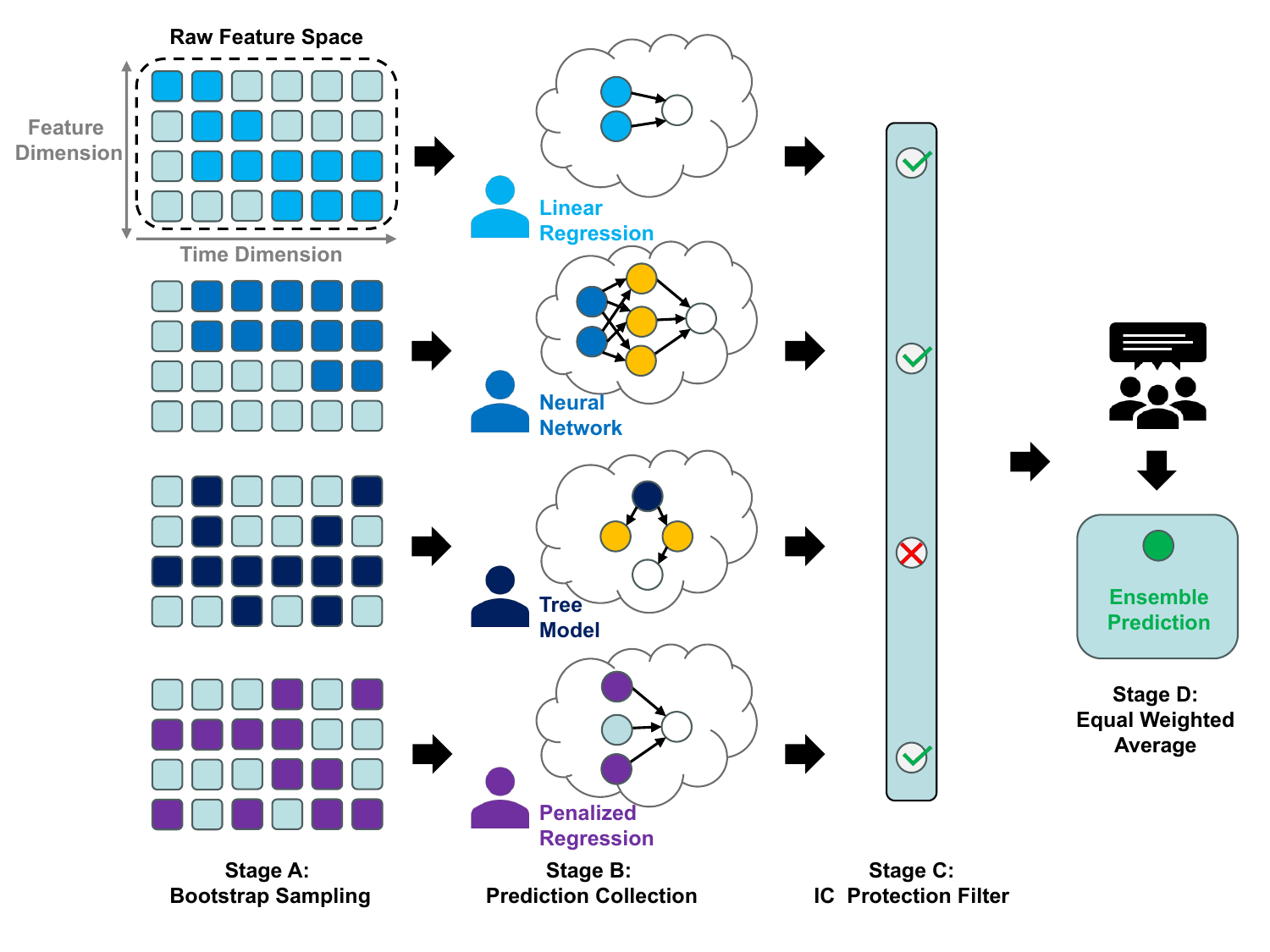}
\caption{Pipeline of the Diversified Expert Ensemble.
\textbf{(A)}~Each predictor receives an independently bootstrap-sampled subset of the raw factor space, combined with a randomly selected feature subset, ensuring diverse training views across the ensemble.
\textbf{(B)}~All $N$ heterogeneous predictors (linear and nonlinear families) produce individual cross-sectional forecasts in parallel.
\textbf{(C)}~An IC protection filter evaluates each predictor on the validation set and removes those whose IC falls below a statistical threshold, eliminating degenerate members while preserving ensemble diversity.
\textbf{(D)}~The surviving predictions are combined via equal-weighted averaging to produce the final ensemble score.}
\Description{A pipeline diagram showing four stages: diverse bootstrap-sampled predictor inputs, parallel forecasts from heterogeneous predictors, validation-based IC filtering, and equal-weighted aggregation of the surviving predictions.}
\label{fig:overview}
\end{figure*}

Building on this observation, we propose \textbf{D}iversified \textbf{E}xpert \textbf{E}nsemble (DEE), a framework that recasts return prediction as the aggregation of heterogeneous expert opinions.
As illustrated in Figure~\ref{fig:overview}, DEE first constructs $N$ predictors drawn from two architecturally distinct families, linear and nonlinear models, each trained on an independently bootstrap-sampled view of the raw factor space with a randomly assigned feature subset.
All predictors then produce individual forecasts in parallel, after which a lightweight quality filter removes members whose validation performance falls below a statistical threshold.
The surviving predictions are combined through equal-weighted averaging, yielding a collective forecast whose accuracy emerges from the complementarity of its constituents rather than the strength of any single member.
Evaluated on the CSI~300 universe with the Alpha158 daily factor set~\citep{yang2020qlibaiorientedquantitativeinvestment}, DEE yields substantial improvements over both conventional neural networks and current state-of-the-art tree-based methods in rank information coefficient and downstream portfolio Sharpe ratio.

Our contributions are summarized as follows:
\begin{itemize}
    \item We formalize the transposition of financial diversification theory from the asset-return space to the prediction-error space, establishing a structural correspondence between portfolio variance decomposition and ensemble forecast error decomposition.
    \item We propose DEE, a heterogeneous bagging framework that combines linear and nonlinear predictors under triple stochastic randomization, achieving state-of-the-art predictive and backtesting performance on CSI~300.
    \item We provide a multi-scale analysis of the ensemble, from individual bias-variance profiles to collective error decomposition, and connect the observed performance gains to the theoretical conditions of collective intelligence, demonstrating that the improvement is structural rather than incidental.
\end{itemize}

\section{Diversified Expert Ensemble}
\label{sec:framework}

\subsection{Problem Setup}

Let $\mathbf{x}_{i,t} \in \mathbb{R}^{F}$ denote the factor vector for stock $i$ at date $t$, and $r_{i,t+1}$ the corresponding forward return.
The goal is to learn a scoring function $f:\mathbb{R}^{F}\to\mathbb{R}$ maximizing the cross-sectional Spearman correlation between $\{f(\mathbf{x}_{i,t})\}$ and $\{r_{i,t+1}\}$ (RankIC).
We operate in a regime where $\mathrm{SNR} \triangleq \mathrm{Var}(g(\mathbf{x}))/\mathrm{Var}(\epsilon) \ll 1$, with $g$ the true conditional mean and $\epsilon$ the idiosyncratic noise.

\subsection{Expert Pool Design}

We instantiate $N$ predictors from two architecturally distinct families, each encoding a different prior over the factor-return mapping.

\paragraph{Linear family.}
Four variants share the affine form $\hat{y}=\mathbf{w}^{\top}\mathbf{x}+b$ but impose different regularization priors: an unregularized baseline, an $\ell_{2}$-penalized variant (uniform shrinkage), an $\ell_{1}$-penalized variant (sparse loadings), and an elastic-net variant ($\ell_{1}+\ell_{2}$, balancing sparsity with stability).
Members with explicit penalties incorporate $\mathcal{L}_{\mathrm{reg}}$ directly into the objective while the optimizer weight decay is disabled, preventing double penalization and preserving the intended diversity of shrinkage profiles.

\paragraph{Nonlinear family.}
A shallow multilayer perceptron with batch normalization captures distributed hierarchical interactions among factors.
A soft decision tree of depth $d$ provides a structurally orthogonal inductive bias: each internal node $n$ routes inputs via $g_{n}(\mathbf{x})=\sigma(\mathbf{a}_{n}^{\top}\mathbf{x}+b_{n})$, and the output is the path-weighted leaf sum $\hat{y}=\sum_{\ell=1}^{2^{d}}P(\ell\mid\mathbf{x})\,\mu_{\ell}$, yielding an axis-aligned piecewise-constant approximation that is fully differentiable.

\subsection{Triple Diversification Protocol}

To minimize the pairwise error covariance in Eq.~\eqref{eq:bvc} (introduced formally in Section~\ref{sec:heterogeneity}), we apply three independent randomization axes:
\begin{enumerate}[leftmargin=1.5em, itemsep=1pt]
    \item \textbf{Bootstrap subsampling.} Each predictor trains on a pool $\mathcal{B}_{k}$ of $\lfloor\beta T\rfloor$ observations drawn with replacement, ensuring no two experts share an identical training distribution.
    \item \textbf{Feature subsampling.} Each predictor observes a random subset $\mathcal{F}_{k}\subset\{1,\dots,F\}$ of size $\lfloor\gamma F\rfloor$, subject to a full-coverage constraint guaranteeing every factor is used by at least one expert.
    \item \textbf{Type assignment.} The $N$ slots are populated by round-robin cycling over the six types, enforcing balanced representation of both families.
\end{enumerate}

\subsection{Aggregation and Quality Control}

The ensemble score is the equal-weight average over active predictors:
\begin{equation}
    \hat{y}_{\mathrm{ens}}(\mathbf{x}) = \frac{1}{|\mathcal{A}|}\sum_{k\in\mathcal{A}} f_{k}(\mathbf{x}_{\mathcal{F}_{k}}),
    \label{eq:agg}
\end{equation}
where $\mathcal{A}$ excludes any predictor whose validation RankIC falls below $\overline{\mathrm{IC}}-\kappa\,\sigma_{\mathrm{IC}}$.
Equal weighting is chosen for its robustness when optimal weight estimation is itself noise-dominated~\citep{demiguel2009optimal}.

\subsection{Training Procedure}

All $N$ predictors are updated within a single synchronous loop (Algorithm~\ref{alg:dee}).
At each step, predictor $k$ performs one gradient update on a mini-batch from $\mathcal{B}_{k}$ restricted to $\mathcal{F}_{k}$.
Early stopping monitors the ensemble-level validation loss, so the collective governs the training horizon rather than any individual member.

\begin{algorithm}[t]
\caption{Diversified Expert Ensemble}
\label{alg:dee}
\begin{algorithmic}[1]
\REQUIRE Data $\{(\mathbf{x}_{t},y_{t})\}_{t=1}^{T}$; predictor count $N$; type cycle $\mathcal{T}$; ratios $\beta,\gamma$; tolerance $\kappa$
\STATE Assign types $t_{k}\!\leftarrow\!\mathcal{T}[k\bmod|\mathcal{T}|]$;\; sample pools $\mathcal{B}_{k}$;\; sample subsets $\mathcal{F}_{k}$
\STATE Initialize $\{f_{k}\}_{k=1}^{N}$ per types and feature dimensions
\FOR{step $=1,\dots,S_{\max}$}
    \FOR{$k=1,\dots,N$}
        \STATE Mini-batch $\mathcal{M}_{k}\!\sim\!\mathcal{B}_{k}$;\; $\mathcal{L}_{k}\!\leftarrow\!\mathcal{L}_{\mathrm{MSE}}+\mathcal{L}_{\mathrm{reg}}^{(k)}$;\; update $f_{k}$
    \ENDFOR
    \IF{validation step}
    \STATE Update early-stopping counter on ensemble loss
\ENDIF
\ENDFOR
\STATE $\mathcal{A}\!\leftarrow\!\{k:\mathrm{IC}_{k}\geq\overline{\mathrm{IC}}-\kappa\,\sigma_{\mathrm{IC}}\}$;\; return Eq.~\eqref{eq:agg}
\end{algorithmic}
\end{algorithm}

\section{Error Complementarity under Heterogeneity}
\label{sec:heterogeneity}

We now explain why mixing linear and nonlinear experts differs from ensembling within a single family.
The analysis starts from the bias-variance-covariance decomposition~\citep{ueda1996generalization, NIPS1994_b8c37e33}:
\begin{equation}
    \mathbb{E}\!\left[\bar{e}^{\,2}\right]
    = \underbrace{\bar{b}^{\,2}}_{\text{mean bias}^{2}}
    + \underbrace{\frac{1}{N}\,\bar{v}}_{\text{mean variance}}
    + \underbrace{\frac{N-1}{N}\,\bar{c}}_{\text{mean covariance}},
    \label{eq:bvc}
\end{equation}
where $\bar{b}$, $\bar{v}$, and $\bar{c}$ denote the average squared bias, average variance, and average pairwise error covariance across the $N$ experts.
Thus, the diversification benefit is governed by $\bar{c}$: the ensemble improves over the mean individual expert only when $\bar{c}<\bar{v}$.

\subsection{Error Structure of Each Family}

Suppose the true conditional mean decomposes as
\[
    g(\mathbf{x}) = g_{L}(\mathbf{x}) + g_{\mathrm{NL}}(\mathbf{x}),
\]
where $g_{L}$ is the best linear projection and $g_{\mathrm{NL}}$ is the residual nonlinear component.
In the low-SNR regime, $\|g_{\mathrm{NL}}\|^{2}\ll\mathrm{Var}(\epsilon)$, yet $g_{\mathrm{NL}}$ is precisely the signal that linear models cannot capture.

\paragraph{Linear expert.}
A linear predictor $f_{L}$ has negligible estimation variance but a systematic bias $b_{L}(\mathbf{x})\approx -g_{\mathrm{NL}}(\mathbf{x})$.
Its error is therefore dominated by the omitted nonlinear signal:
\begin{equation}
    e_{L}(\mathbf{x}) \;\approx\; -g_{\mathrm{NL}}(\mathbf{x}) + \eta_{L},
    \label{eq:err_linear}
\end{equation}
where $\eta_{L}$ is a small estimation noise term with $\mathrm{Var}(\eta_{L})=O(\sigma^{2}/T)$.

\paragraph{Nonlinear expert.}
A nonlinear predictor $f_{\mathrm{NL}}$ has sufficient capacity to approximate $g$, so its bias is small.
However, in the low-SNR regime it partially fits the noise $\epsilon$, inflating its variance:
\begin{equation}
    e_{\mathrm{NL}}(\mathbf{x}) \;\approx\; \eta_{\mathrm{NL}},
    \qquad
    \mathrm{Var}(\eta_{\mathrm{NL}})
    = O\!\left(\frac{p_{\mathrm{NL}}}{T}\,\sigma^{2}\right),
    \label{eq:err_nn}
\end{equation}
where $p_{\mathrm{NL}}$ is the effective parameter count of the network.

\subsection{Covariance Suppression}

The cross-covariance between a linear and a nonlinear expert is
\begin{equation}
    \mathrm{Cov}(e_{L},\,e_{\mathrm{NL}})
    =
    \mathrm{Cov}(-g_{\mathrm{NL}}+\eta_{L},\;\eta_{\mathrm{NL}})
    =
    -\mathrm{Cov}(g_{\mathrm{NL}},\,\eta_{\mathrm{NL}})
    +
    \mathrm{Cov}(\eta_{L},\,\eta_{\mathrm{NL}}).
    \label{eq:cov_cross}
\end{equation}
Two mechanisms drive both terms toward zero:
\begin{enumerate}[leftmargin=1.5em, itemsep=2pt]
    \item \textbf{Stochastic independence.} When $f_{L}$ and $f_{\mathrm{NL}}$ are trained on different bootstrap pools and feature subsets, $\eta_{L}$ and $\eta_{\mathrm{NL}}$ depend on largely disjoint data, so $\mathrm{Cov}(\eta_{L},\eta_{\mathrm{NL}})\to 0$.
    \item \textbf{Signal orthogonality.} The term $\mathrm{Cov}(g_{\mathrm{NL}},\eta_{\mathrm{NL}})$ measures whether the network's estimation noise is correlated with the true nonlinear signal. Under independent training samples this covariance is $O(1/T)$ and vanishes asymptotically.
\end{enumerate}

In contrast, two linear experts $f_{L}^{(1)},f_{L}^{(2)}$ share the same structural bias $-g_{\mathrm{NL}}$, giving
\begin{equation}
    \mathrm{Cov}(e_{L}^{(1)},e_{L}^{(2)})
    \;\approx\;
    \mathrm{Var}(g_{\mathrm{NL}})
    +
    \mathrm{Cov}(\eta_{L}^{(1)},\eta_{L}^{(2)})
    \;\geq\;
    \mathrm{Var}(g_{\mathrm{NL}}).
    \label{eq:cov_same}
\end{equation}
This covariance is bounded below by the nonlinear signal variance, regardless of randomization.
An analogous argument applies to two nonlinear experts that share similar noise-fitting patterns.

\begin{proposition}[Heterogeneity advantage]
\label{prop:hetero}
Let $\bar{c}_{\mathrm{homo}}$ and $\bar{c}_{\mathrm{hetero}}$ denote the mean pairwise error covariance of a homogeneous ensemble and a heterogeneous ensemble, respectively, both of size $N$ and trained under identical bootstrap and feature-subsample protocols.
Under the error model in Eqs.~\eqref{eq:err_linear}--\eqref{eq:err_nn},
\[
    \bar{c}_{\mathrm{hetero}}
    \leq
    \bar{c}_{\mathrm{homo}}
    -
    \mathrm{Var}(g_{\mathrm{NL}})
    +
    O(1/T),
\]
and the inequality is strict whenever $g_{\mathrm{NL}}\not\equiv 0$.
\end{proposition}

\begin{proof}[Proof sketch]
For a homogeneous linear ensemble, every pair shares the bias component $-g_{\mathrm{NL}}$, contributing $\mathrm{Var}(g_{\mathrm{NL}})$ to every pairwise covariance by Eq.~\eqref{eq:cov_same}.
For a heterogeneous pair $(f_{L},f_{\mathrm{NL}})$, this shared component is absent by Eq.~\eqref{eq:cov_cross}.
Averaging over all $\binom{N}{2}$ pairs, and noting that a balanced ensemble contains $\Theta(N^{2})$ cross-family pairs, yields the stated bound.
\end{proof}

Thus, homogeneous ensembles carry an irreducible covariance floor that heterogeneous ensembles can remove.

\begin{figure}[t]
    \centering
    \includegraphics[width=0.48\textwidth]{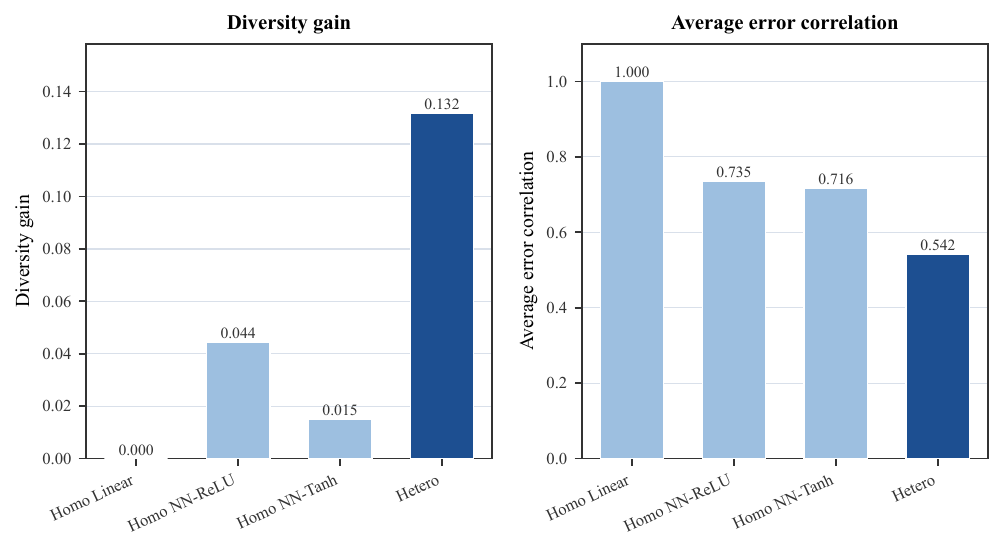}
    \caption{
    Synthetic validation of error complementarity.
    Homogeneous ensembles use a single expert family, while the heterogeneous ensemble mixes linear, ReLU MLP, and Tanh MLP experts.
    Errors are computed against the noiseless target.
    The heterogeneous ensemble exhibits a larger diversity gain and a lower average pairwise error correlation, supporting Proposition~\ref{prop:hetero}.
    }
    \Description{A plot comparing homogeneous and heterogeneous ensembles on diversity gain and average pairwise error correlation; the heterogeneous ensemble has greater diversity gain and lower error correlation.}
    \label{fig:sim_complementarity}
\end{figure}

\paragraph{Controlled validation.}
We simulate a regression target containing three components: a linear term, a smooth interaction term, and a localized nonlinear bump.
We train three expert families: linear regression, a ReLU MLP, and a Tanh MLP.
A homogeneous ensemble uses only one family, whereas the heterogeneous ensemble cycles through the three families.
To avoid common observation noise inflating all error correlations, we compute approximation errors $e_k=f_k-g$ against the noiseless target.
For each ensemble, we report
\[
    \text{diversity gain}
    =
    \overline{\mathrm{MSE}}
    -
    \mathrm{MSE}_{\mathrm{ens}},
\]
and the average pairwise error correlation
\[
    \frac{1}{\binom{N}{2}}
    \sum_{i<j}
    \mathrm{Corr}(e_i,e_j).
\]
Figure~\ref{fig:sim_complementarity} shows that the heterogeneous ensemble achieves both a larger diversity gain and a lower average error correlation than the homogeneous ensembles, consistent with Proposition~\ref{prop:hetero}.

\section{Diversification Gain under Low Signal}
\label{sec:scaling}

Section~\ref{sec:heterogeneity} showed that heterogeneity suppresses $\bar{c}$.
We now ask: given a fixed residual correlation $\rho$, how does the ensemble benefit scale with $N$, and why is this benefit amplified in the low-SNR regime?

\subsection{Variance Decay and the Markowitz Correspondence}

Assume $N$ experts with common error variance $\sigma^{2}$ and common pairwise correlation $\rho\in[0,1)$.
The variance of the averaged error is
\begin{equation}
    \mathrm{Var}(\bar{e})
    =
    \sigma^{2}\!\left(\rho + \frac{1-\rho}{N}\right).
    \label{eq:var_decay}
\end{equation}
This expression is structurally identical to the variance of an equally weighted portfolio of $N$ assets with pairwise return correlation $\rho$~\citep{e5a1bb8f-41b7-35c6-95cd-8b366d3e99bc}.
The marginal reduction from adding the $(N{+}1)$-th expert is
\begin{equation}
    \Delta_{N}
    \triangleq
    \mathrm{Var}(\bar{e}_{N})
    -
    \mathrm{Var}(\bar{e}_{N+1})
    =
    \frac{\sigma^{2}(1-\rho)}{N(N+1)}.
    \label{eq:marginal}
\end{equation}
Hence, whenever $\rho<1$, each additional expert reduces variance, but the marginal gain decreases as $N$ grows.

\subsection{Effective SNR Amplification}

Let $S=\mathrm{Var}(g)$ denote the signal variance and $V$ the idiosyncratic noise variance.
The individual expert's signal-to-noise ratio is
\[
    \mathrm{SNR} = \frac{S}{V},
\]
and its population $R^{2}$ is
\[
    R^{2}_{1}
    =
    \frac{S}{S+V}
    =
    \frac{\mathrm{SNR}}{1+\mathrm{SNR}}.
\]
In the low-SNR regime, $R^{2}_{1}\approx \mathrm{SNR}\ll 1$.

From Eq.~\eqref{eq:var_decay}, averaging $N$ experts reduces the idiosyncratic noise variance from $V$ to
\[
    V\left(\rho + \frac{1-\rho}{N}\right).
\]
When heterogeneity drives $\rho\approx 0$, this becomes $V/N$.
Therefore, the ensemble $R^{2}$ satisfies
\begin{equation}
    R^{2}_{N}
    =
    \frac{S}{S+V/N}
    =
    \frac{N\cdot\mathrm{SNR}}{1+N\cdot\mathrm{SNR}}.
    \label{eq:snr_amp}
\end{equation}

\begin{proposition}[SNR amplification]
\label{prop:snr}
Under the conditions of Eq.~\eqref{eq:snr_amp}, the ensemble effectively operates at an amplified signal-to-noise ratio
\[
    \mathrm{SNR}_{\mathrm{eff}} = N\cdot\mathrm{SNR}.
\]
For example, if $\mathrm{SNR}=0.02$ and $N=30$, then
\[
    R^{2}_{N}
    \approx
    \frac{0.6}{1.6}
    \approx
    0.38,
\]
whereas $R^{2}_{1}\approx 0.02$.
\end{proposition}

This reveals a key asymmetry between high-SNR and low-SNR domains.
In high-SNR tasks, a single strong model may already achieve $R^{2}\approx 1$, leaving little room for ensemble improvement.
In low-SNR financial prediction, each individual expert is weak, but averaging many weakly correlated experts can substantially amplify the effective signal.

\begin{figure}[H]
    \centering
    \includegraphics[width=0.48\textwidth]{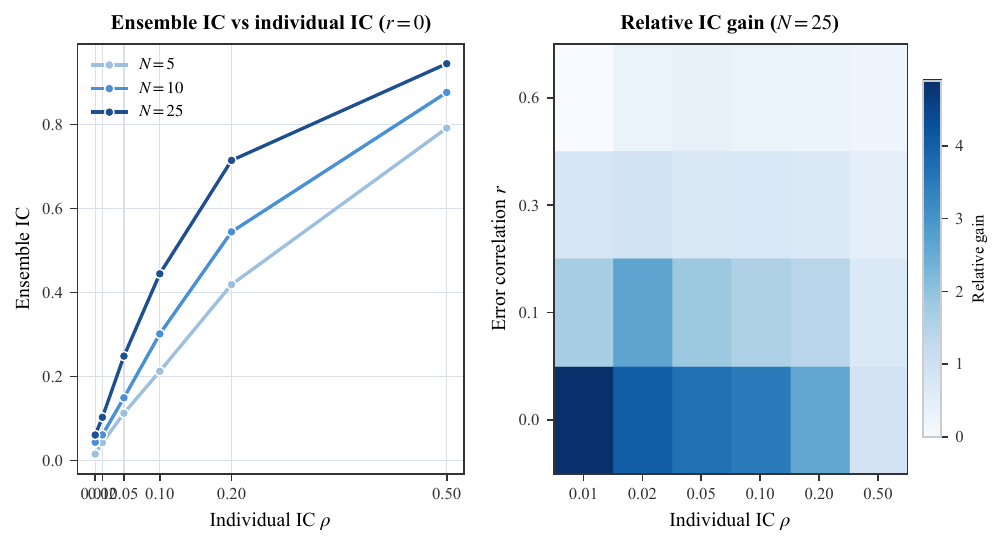}
    \caption{
    Synthetic validation of diversification gain under low signal.
    Predictors are generated with prescribed individual IC $\rho$ and pairwise error correlation $r$.
    Left: ensemble IC versus individual IC for different ensemble sizes.
    Right: relative IC gain as a function of $\rho$ and $r$.
    The gain is largest when individual signal is weak and errors are weakly correlated, consistent with Proposition~\ref{prop:snr}.
    }
    \Description{Two synthetic plots showing that diversification gains are largest for weak individual signals and weakly correlated prediction errors, and increase with ensemble size.}
    \label{fig:sim_low_signal}
\end{figure}

\subsection{Interaction of Heterogeneity and Scale}

The two mechanisms are multiplicative rather than additive.
Heterogeneity drives $\rho\to 0$, which is the precondition for the $1/N$ variance decay in Eq.~\eqref{eq:var_decay}.
Increasing $N$ then amplifies the SNR, making the bias term $\bar{b}^{\,2}$ in Eq.~\eqref{eq:bvc} the dominant residual error and motivating the inclusion of nonlinear experts to reduce $\bar{b}$.

\begin{remark}[Design implication]
\label{rem:design}
The analysis prescribes a two-pronged design:
(i) maximize heterogeneity to minimize $\rho$;
(ii) increase $N$ to exploit the resulting $1/N$ decay.
Our triple diversification protocol addresses (i), while the round-robin type assignment and scalable training loop address (ii).
The IC-based quality filter ensures that increasing $N$ does not introduce harmful members that would elevate $\rho$.
\end{remark}

\paragraph{Controlled validation.}
We generate predictors with prescribed individual IC $\rho$ and pairwise error correlation $r$:
\begin{equation}
    p_k
    =
    \rho\, y
    +
    \sqrt{1-\rho^{2}}
    \left(
        \sqrt{r}\,z_0
        +
        \sqrt{1-r}\,z_k
    \right),
    \label{eq:sim_ic}
\end{equation}
where $y$, $z_0$, and $z_k$ are independent standard Gaussian variables.
The common shock $z_0$ induces correlated errors, while $z_k$ produces idiosyncratic errors.
We form the equal-weight ensemble $\bar{p}=N^{-1}\sum_{k=1}^{N}p_k$ and measure its IC with $y$.
Figure~\ref{fig:sim_low_signal} reports two results.
First, when the individual IC $\rho$ is small, increasing $N$ substantially raises the ensemble IC.
Second, the relative gain is largest for small $\rho$ and small $r$, and diminishes as error correlation increases.
This behavior is consistent with Eqs.~\eqref{eq:var_decay} and~\eqref{eq:snr_amp}, and with the residual-correlation floor discussed below.

\subsection{Finite-Sample Considerations}

In practice, $\rho$ does not reach exactly zero due to shared factor structure and finite training data.
Let $\rho_{\mathrm{res}}>0$ denote the residual correlation after diversification.
The asymptotic variance floor is then $\sigma^{2}\rho_{\mathrm{res}}$, and the effective ensemble size saturates at
\[
    N^{*}
    \approx
    \frac{1-\rho_{\mathrm{res}}}{\rho_{\mathrm{res}}}.
\]
For $\rho_{\mathrm{res}}=0.05$, $N^{*}\approx 19$, indicating that most of the diversification benefit is captured with $N=20$ to $30$ experts, consistent with our empirical findings.

\section{Related Work}

\subsection{Predictive Models for Asset Pricing}

Factor-based return prediction has evolved from linear specifications to increasingly complex nonlinear architectures.
The foundational paradigm, exemplified by the Fama-French multi-factor model~\citep{fama1993common} and commercial risk models such as Barra~\citep{rosenberg1974extra}, relies on linear regression to estimate factor exposures, offering interpretability and statistical stability but limited capacity for interaction effects.
Neural network approaches, including multilayer perceptrons~\citep{kimoto1990stock}, recurrent architectures~\citep{fischer2018deep}, and attention-based models~\citep{liu2020adaptive, yang2020qlibaiorientedquantitativeinvestment}, have demonstrated improved in-sample fit on historical factor data.
However, in the low signal-to-noise regime characteristic of daily equity returns, these models frequently overfit transient noise patterns, yielding degraded out-of-sample performance~\citep{doi:https://doi.org/10.1002/9780470061602.eqf15018, gu2020empirical}.
Gradient-boosted decision trees, particularly XGBoost~\citep{10.1145/2939672.2939785} and LightGBM~\citep{NIPS2017_6449f44a}, currently achieve the strongest empirical results on standard benchmarks~\citep{gu2020empirical}, yet they constitute a single function class refined through sequential residual fitting, providing no explicit mechanism for incorporating architecturally diverse inductive biases.
The present work departs from this trajectory by treating model heterogeneity not as a limitation to be overcome but as a resource to be exploited.

\subsection{Ensemble Strategies in Financial Prediction}

Ensemble learning combines multiple predictors to reduce generalization error.
Bagging~\citep{breiman1996bagging} and random forests~\citep{breiman2001random} introduce diversity through bootstrap subsampling and feature randomization, while stacking~\citep{wolpert1992stacked} learns a meta-level combiner over base predictions.
In quantitative finance, Double Ensemble~\citep{9338413} extends boosting by jointly reweighting samples and features across iterations, and snapshot ensembles~\citep{huang2017snapshotensemblestrain1} aggregate multiple local minima of a single network's training trajectory.
These methods have demonstrated consistent gains over individual models on financial datasets.
Nevertheless, existing financial ensembles are predominantly homogeneous, composing multiple instances of the same architecture family, typically tree-based learners.
The resulting inter-member correlation remains elevated, limiting the achievable variance reduction.
Our framework explicitly constructs heterogeneity by combining linear and neural families under a unified bagging protocol, targeting the decorrelation that homogeneous ensembles cannot attain.

\subsection{Diversification Theory and Error Decomposition}

The mathematical foundation for combining imperfect estimates is well established in both finance and statistical learning theory.
Markowitz's mean-variance framework~\citep{e5a1bb8f-41b7-35c6-95cd-8b366d3e99bc} demonstrates that portfolio risk decreases as the correlation among constituent assets declines, a principle that has governed portfolio construction for over seven decades.
In parallel, the bias-variance decomposition~\citep{geman1992neural} and its extension to ensembles via the bias-variance-covariance formulation~\citep{ueda1996generalization, NIPS1994_b8c37e33} reveal that the expected squared error of an averaged predictor depends critically on the covariance among member errors.
Breiman~\citep{breiman1996bagging} formalized this connection, showing that the ensemble gain is governed by the average pairwise correlation of base learners.
Kuncheva and Whitaker~\citep{kuncheva2003measures} further catalogued diversity measures and their relationship to ensemble accuracy.
Despite this rich theoretical apparatus, the structural isomorphism between portfolio variance reduction and ensemble error reduction has not been explicitly leveraged as a design principle for financial prediction systems.
We make this correspondence operational, using it to guide both the construction and the analysis of our expert ensemble.

\subsection{Collective Intelligence}

Beyond the mathematical decomposition, the theory of collective intelligence provides a complementary framework. Extensions of the Condorcet jury theorem establish that aggregating signals from independent and competent agents converges to the optimal decision as group size increases~\citep{austen1996information, nitzan1982optimal}. This principle is empirically validated across early statistical demonstrations~\citep{galton1907vox, armstrong1989combining}, modern prediction markets~\citep{wolfers2004jep}, and decentralized complex systems~\citep{arrow2008promise}. Furthermore, foundational research formally proves that prediction accuracy critically relies on the diversity and independence of individual models~\citep{NIPS1994_b8c37e33, hong2004groups}. While these theoretical conditions map naturally onto predictive ensembles, their application remains predominantly confined to classification tasks~\citep{dietterich2000ensemble}. We extend this reasoning to continuous financial return prediction by demonstrating that architectural heterogeneity and stochastic training structurally satisfy the necessary conditions. Consequently, the resulting collective exhibits emergent predictive capabilities exceeding those of any individual member.

\section{Experiments}
\label{sec:experiments}

\subsection{Experimental Setup}

\paragraph{Dataset.}
We evaluate on the CSI~300 constituent universe using the Alpha158 daily factor set provided by the Qlib quantitative investment platform~\citep{yang2020qlibaiorientedquantitativeinvestment}.
Alpha158 comprises 158 carefully engineered price-volume factors spanning multiple lookback windows.
Following the standard temporal split, we use 2008-01-01 through 2014-12-31 for training, 2015-01-01 through 2016-12-31 for validation, and 2017-01-01 through 2020-12-31 for testing.
All experiments are repeated 10 times with different random seeds; we report mean and standard deviation.

\paragraph{Baselines.}
We compare against six representative methods covering the major model families in quantitative prediction:
Linear Regression (linear baseline),
MLP (nonlinear neural baseline),
Random Forest~\citep{breiman2001random} (bagged tree ensemble),
XGBoost~\citep{10.1145/2939672.2939785} and LightGBM~\citep{NIPS2017_6449f44a} (gradient-boosted trees, current empirical SOTA),
and Double Ensemble~\citep{9338413} (sample- and feature-reweighted tree ensemble).
We exclude sequence models such as GRU~\citep{cho-etal-2014-learning} and ALSTM~\citep{ijcai2017p366} because they require an external feature selection step reducing the 158 factors to approximately 20 inputs for temporal modeling, making the comparison non-uniform.

\paragraph{Evaluation metrics.}
Predictive quality is measured by the information coefficient (IC, Pearson correlation between predicted scores and forward returns) and rank IC (Spearman rank correlation), both computed cross-sectionally per trading day and averaged over the test period.
Portfolio performance is assessed via a vectorized top-$K$ backtest: each day we form an equal-weight long portfolio of the $K=50$ stocks with the highest predicted scores and compute annualized return, annualized Sharpe ratio, and maximum drawdown.

\paragraph{DEE configuration.}
Unless otherwise stated, DEE uses $N=50$ predictors allocated as 10 linear, 2 ridge, 5 lasso, 3 elastic net, 15 MLP (hidden width 32), and 15 soft tree (depth 12), with regularization strength $\alpha=10^{-3}$ for all penalized linear variants.
Each predictor trains on a bootstrap pool ($\beta=0.8789$) restricted to a random feature subset ($\gamma=0.5$), optimized via Adam (lr $10^{-3}$, weight decay $10^{-4}$, batch size 8192, MSE loss) for up to 8000 steps with early stopping (patience 50, evaluated every 20 steps).
The IC filter tolerance is $\kappa=1.5$; no dropout is applied.

\subsection{Predictive Performance}
\label{sec:exp_predictive}

Table~\ref{tab:predictive} reports predictive metrics on the CSI~300 test set.
DEE achieves the highest IC (0.048) and RankIC (0.060), surpassing XGBoost by 4.3\% and 1.7\% in RankIC, and MLP by 13.2\%.
The improvement over Double Ensemble, the strongest ensemble baseline, is 7.1\% in RankIC.
Notably, the gap between DEE and the tree-based SOTA is achieved without any tree component in the nonlinear family contributing gradient-boosted structure; the soft decision trees in DEE are bounded-depth weak learners serving purely as decorrelation devices.
The consistently low standard deviations across all methods (below 0.002) confirm the statistical reliability of the reported rankings.

\begin{table}[H]
\centering
\caption{Predictive performance on CSI~300 (Alpha158, daily, 2017--2020). Best in \textbf{bold}, second best \underline{underlined}. Values are mean $\pm$ std over 10 runs; $\dagger$ denotes a single run.}
\label{tab:predictive}
\small
\begin{tabular}{l cc}
\toprule
\textbf{Model} & \textbf{IC} & \textbf{RankIC} \\
\midrule
Linear Reg.$^{\dagger}$ & 0.033 & 0.044 \\
MLP             & 0.040 $\pm$ 0.002 & 0.053 $\pm$ 0.001 \\
Random Forest   & 0.044 $\pm$ 0.000 & 0.054 $\pm$ 0.001 \\
Double Ensemble & 0.044 $\pm$ 0.001 & 0.056 $\pm$ 0.001 \\
LightGBM        & 0.045 $\pm$ 0.001 & 0.058 $\pm$ 0.000 \\
XGBoost         & \underline{0.046} $\pm$ 0.000 & \underline{0.059} $\pm$ 0.001 \\
\midrule
DEE (Ours)      & \textbf{0.048} $\pm$ 0.001 & \textbf{0.060} $\pm$ 0.001 \\
\bottomrule
\end{tabular}
\end{table}

\subsection{Ablation Studies}
\label{sec:exp_ablation}

\paragraph{Effect of architectural heterogeneity.}
Table~\ref{tab:ablation_hetero} isolates the contribution of type diversity by fixing $N=50$ and varying the predictor family.
The full six-type DEE outperforms the three-type variant (linear, MLP, tree) by 2.2\% in RankIC, confirming that the additional regularization variants within the linear family contribute meaningful decorrelation.
More strikingly, any single-family ensemble underperforms the heterogeneous combination: MLP-only and tree-only ensembles achieve RankIC of 0.055 and 0.054 respectively, while the linear-only ensemble reaches only 0.051.
The gap between the full DEE (0.060) and the best single-family ensemble (MLP, 0.055) is 9.1\%, providing direct empirical support for Proposition~\ref{prop:hetero}: heterogeneous ensembles eliminate the shared-bias covariance floor that constrains homogeneous ones.

\begin{table}[t]
\centering
\caption{Ablation on predictor family composition ($N=50$, CSI~300). Mean $\pm$ std over 10 runs.}
\label{tab:ablation_hetero}
\small
\begin{tabular}{l cc}
\toprule
\textbf{Predictor Family} & \textbf{IC} & \textbf{RankIC} \\
\midrule
Linear only           & 0.039 $\pm$ 0.000 & 0.051 $\pm$ 0.000 \\
Tree only             & 0.041 $\pm$ 0.001 & 0.054 $\pm$ 0.001 \\
MLP only              & 0.043 $\pm$ 0.001 & 0.055 $\pm$ 0.001 \\
Linear + MLP + Tree   & \underline{0.046} $\pm$ 0.001 & \underline{0.059} $\pm$ 0.001 \\
\midrule
Full DEE (6 types)    & \textbf{0.048} $\pm$ 0.001 & \textbf{0.060} $\pm$ 0.001 \\
\bottomrule
\end{tabular}
\end{table}

\begin{figure}[t]
    \centering
    \includegraphics[width=\linewidth]{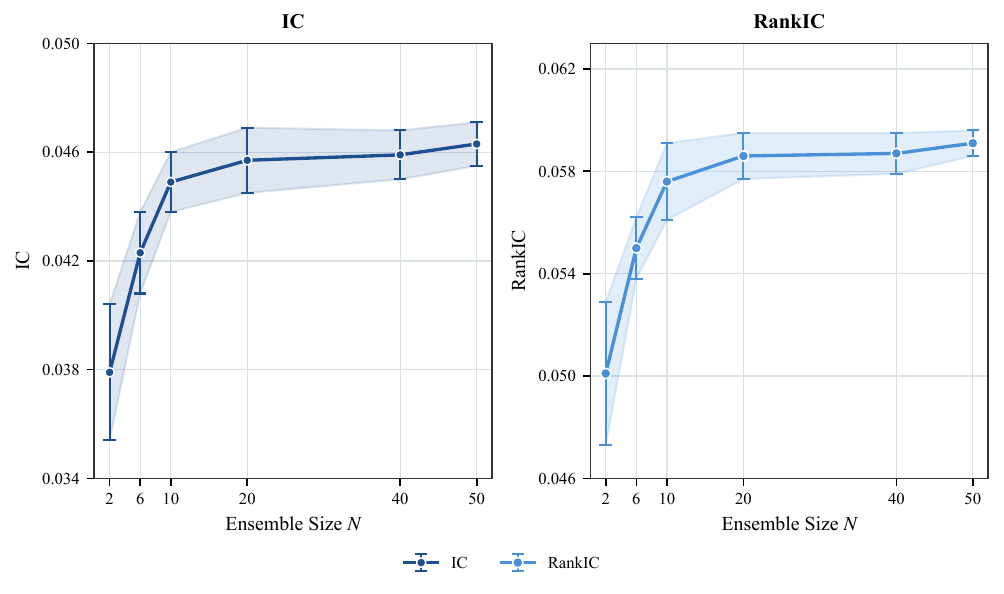}
    \caption{Predictive performance versus ensemble size $N$
      (1:1 linear:MLP, CSI~300). Shaded regions denote $\pm1$ std over
      10 runs. Both metrics exhibit monotonic improvement with diminishing
      marginal returns, consistent with the $\mathcal{O}(1/N)$
      diversification decay of Eq.~\eqref{eq:var_decay}.}
    \Description{A line plot of IC and RankIC versus ensemble size from 2 to 50, showing monotonic improvement with diminishing marginal returns and narrowing uncertainty bands.}
    \label{fig:scaling}
\end{figure}

\paragraph{Effect of ensemble size.}
Figure~\ref{fig:scaling} traces predictive performance as $N$ increases from 2 to 50, holding the type ratio fixed at 1:1 (linear:MLP).
Both IC and RankIC improve monotonically, with the marginal gain shrinking as $N$ grows: the RankIC increment from $N=2$ to $N=10$ is 0.0075, while that from $N=20$ to $N=50$ is only 0.0005.
This pattern matches the $\mathcal{O}(1/N)$ variance decay predicted by Eq.~\eqref{eq:var_decay}.
Concurrently, the shaded confidence band narrows from $\pm$0.003 at $N=2$ to $\pm$0.001 at $N=50$, confirming that larger ensembles not only improve mean performance but also stabilize it across random seeds, consistent with the diversification interpretation of Section~\ref{sec:scaling}.
The curve begins to plateau around $N=20$, aligning with the saturation estimate $N^{*}\approx 19$ derived from the residual correlation analysis in Section~\ref{sec:scaling}.

\subsection{Portfolio Backtest Performance}
\label{sec:exp_backtest}

Table~\ref{tab:backtest} translates predictive signals into portfolio-level outcomes via the top-50 equal-weight strategy.
DEE achieves the highest annualized return (38.56\%) and Sharpe ratio (1.846), while simultaneously recording the second lowest maximum drawdown (17.15\%) among all methods.
Relative to XGBoost, DEE improves the Sharpe ratio by 5.1\% and reduces drawdown by 1.7 percentage points.
The contrast with MLP is more pronounced: DEE improves annualized return by 6.1 percentage points and Sharpe by 13.8\%, confirming that the ensemble framework effectively remedies the overfitting pathology of standalone neural networks in low-SNR regimes.
Linear Regression, lacking any nonlinear capacity, produces the weakest portfolio across all three metrics, underscoring the necessity of nonlinear experts even in a predominantly linear factor environment.

The co-occurrence of highest return and lowest drawdown is noteworthy.
Individual tree models (XGBoost, LightGBM, Double Ensemble) cluster around 36\% annual return with drawdowns near 19\%, suggesting that their correlated error structures expose the portfolio to common tail events.
DEE's diversified error profile, by contrast, disperses this exposure, yielding a more favorable return-drawdown trade-off that echoes the risk reduction mechanism of a well-diversified asset portfolio.

\begin{table}[t]
\centering
\caption{Portfolio backtest on CSI~300 (top-50 equal-weight, daily rebalancing, 2017--2020). Mean $\pm$ std over 10 runs; $\dagger$ denotes a single run. MDD = maximum drawdown (lower is better).}
\label{tab:backtest}
\small
\begin{tabular}{l ccc}
\toprule
\textbf{Model} & \textbf{Ann. Return (\%)} & \textbf{Sharpe Ratio} & \textbf{Max DD (\%)} \\
\midrule
Linear Reg.$^{\dagger}$ & 25.32 & 1.214 & 26.25 \\
MLP             & 32.42 $\pm$ 1.344 & 1.622 $\pm$ 0.054 & \textbf{16.94} $\pm$ 1.422 \\
Random Forest   & 35.75 $\pm$ 0.738 & 1.688 $\pm$ 0.036 & 17.98 $\pm$ 1.109 \\
Double Ensemble & 36.10 $\pm$ 0.823 & 1.742 $\pm$ 0.036 & 19.21 $\pm$ 0.946 \\
LightGBM        & 36.32 $\pm$ 1.094 & 1.745 $\pm$ 0.055 & 19.33 $\pm$ 1.243 \\
XGBoost         & \underline{36.33} $\pm$ 0.716 & \underline{1.757} $\pm$ 0.034 & 18.82 $\pm$ 0.662 \\
\midrule
DEE (Ours)      & \textbf{38.56} $\pm$ 0.871 & \textbf{1.846} $\pm$ 0.042 & \underline{17.15} $\pm$ 0.923 \\
\bottomrule
\end{tabular}
\end{table}

\begin{figure}[t]
\centering
\includegraphics[width=\linewidth]{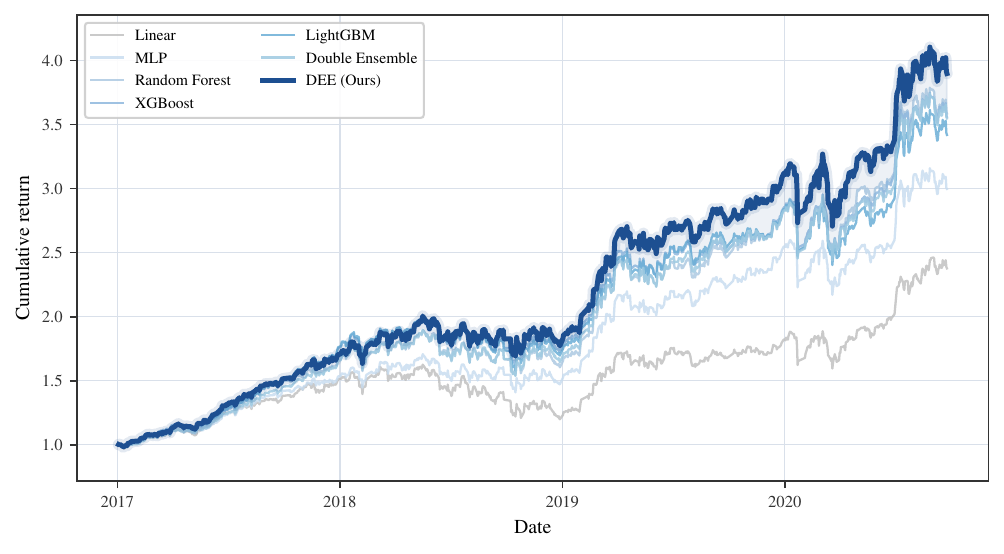}
\caption{Cumulative return of the top-50 equal-weight portfolio over the
  CSI~300 test period (2017-01 to 2020-12).}
\Description{A line plot of cumulative returns for the top-50 equal-weight portfolio during the CSI 300 test period from January 2017 through December 2020.}
\label{fig:cumret}
\end{figure}

Figure~\ref{fig:cumret} plots the cumulative return trajectories over the
test period. DEE attains a competitive cumulative return, performing on par
with the strongest baselines across the full horizon, with a broadly stable
relative positioning under the differing market conditions within the window.

\section{Conclusion}

In this work, we presented a unified study of ensemble learning for financial prediction, grounded in the interplay between error
complementarity and diversification. Across theory and experiments, we found that predictive gains are governed less by the strength of
individual models than by the decorrelation of their errors, and that heterogeneity is the principal lever for unlocking such gains. The
resulting diversification law predicts a diminishing marginal return that decays as $\mathcal{O}(1/N)$, offering a principled account of when additional capacity ceases to help.

Despite these insights, the present study has notable limitations. The improvements over strong baselines, while consistent, remain bounded in magnitude, reflecting the fact that variance reduction cannot compensate for fundamentally weak signal. Our analysis also provides only a limited view of robustness: the behavior of ensembles under regime shifts,
distributional stress, and realistic frictions has not been systematically characterized, and the theory rests on simplified correlation assumptions that abstract away market microstructure.

Several directions follow naturally. A deeper understanding of how heterogeneity gives rise to emergent collaboration within an ensemble could inform a principled design of model composition and interaction, moving beyond ad hoc mixing toward ensembles that are diverse by construction. Equally important is robustness-oriented design, including regime-aware and uncertainty-driven strategies that preserve
diversification benefits under non-stationary conditions. Together, these avenues point toward ensemble systems that are not only more accurate but more reliable in the low-signal, high-noise environments that define financial markets.

\begin{acks}
The author sincerely thanks Chao Zhang for his valuable guidance and insightful discussions throughout this work.
\end{acks}

\bibliographystyle{ACM-Reference-Format}
\bibliography{software}


\end{document}